# States, Modes, Fields, and Photons in Quantum Optics

*Michael G. Raymer[1*] and Paul Polakos[2]*

[1.] Department of Physics, and Oregon Center for Optical, Molecular and Quantum Science, University of Oregon, Eugene, OR 97403, USA
[2.] Cisco Systems, New York, NY 10119, USA

* Corresponding author: raymer@uoregon.edu

**Abstract**

The quantum nature of light enables potentially revolutionary communication technologies. Key to advancing this area of research is a clear understanding of the concepts of states, modes, fields, and photons. The concept of field modes carries over from classical optics, while the concept of state has to be considered carefully when treating light quantum mechanically. The term 'photon' is an overloaded identifier in the sense that it is often used to refer to either a quantum particle or the state of a field. This overloading, often used without placing in context, has the potential to obfuscate the physical processes that describe the reality we measure. We review the uses and relationships between these concepts using modern quantum optics theory, including the concept of a photon wave function, the modern history of which was moved forward in a groundbreaking paper in this journal by Iwo Białynicki-Birula, to whom this article is dedicated.

## 1. Introduction

When beginning the study of quantum optics it is natural to ask, "What is a photon?" But perhaps a better question is, "What is a quantum field?" Given that quantum theory is agnostic to the names we give to the mathematical elements of the theory, when does it matter how we name and interpret them? Properly conceptualizing and naming the elements of theory helps when trying to build intuition about a problem without the benefit of having a complete mathematical solution at hand. This contribution to the Special Issue dedicated to Professor Iwo Białynicki-Birula reviews in a tutorial manner the role of states, modes, fields, and photons in quantum optics, recognizing his important contributions to the subject.[i] We hope to enlighten researchers who are perhaps new to the field, such as those working in the classical networks arena and now starting to consider the potentially useful applications of quantum networks. We review the concept of a photon wave function, the modern history of which begins more-or-less with a paper in this journal by Białynicki-Birula [1] and a contemporaneous paper by John Sipe [2].

States, modes, and fields are concepts that apply to both classical and quantum domains. The paper reviews in a pedagogical style how these concepts arise and are defined within the two domains, describes how quantization of EM field excitations introduces new (and measurable) behaviors, and clarifies the connections between the two domains.

---

[i] Acta Phys. Pol. A 143, S28 (2023); Doi: 10.12693/APhysPolA.143.S28

In the domain of applications, we note that in any quantum optical computing or communication system it is required to control the states of light that interact to carry out a quantum information processing (QIP) task. If control is imprecise, 'errors' can occur. In fact, such errors are a main barrier to developing scalable QIP. [3] While for single-particle qubits (e.g. spin of an electron) the concept of state is clear and routine, for optical qubits such is not the case due to the multimode nature of the electromagnetic field, and it is worthwhile discussing some of the subtleties that arise in this case. Much has been written on the general subject, such as [4, 5, 6,7, 8] and studies cited therein, but we aim here to cover topics not widely emphasized, in particular the quantization of the EM field in terms of temporal (wave-packet) modes.

**2. Classical Fields, Modes, and States**

In the classical-physics description, light is a transverse wave of the electromagnetic or EM field (electric and magnetic fields propagating together). Maxwell's equations provide a means to calculate the energy per unit volume stored in the EM field, which may vary continuously. They also provide a wave equation that allow us to calculate the temporal and spatial evolution of the EM field, transporting energy and momentum.

A monochromatic plane wave in free space is identified by specifying values for four distinct attributes (degrees of freedom), any of which can be used to encode information: polarization and three spatial propagation constants $k_x, k_y, k_z$. In a beamlike geometry it is often useful to restate these degrees of freedom as polarization, two spatial degrees of freedom describing the transverse beam profile, and frequency, $\omega = c(k_x^2 + k_y^2 + k_z^2)^{1/2}$, where c is the speed of light. In either case, the four degrees of freedom define a 'mode' of the EM field. A mode can be thought of as a 'container,' into which differing amounts of energy and momentum can be deposited and carried along by the wave.

**Definition**: A *classical electromagnetic field* is a physical entity of infinite spatial extent that can transport energy and momentum in the form of wave-like excitations.

**Definition**: A *mode* $\mathbf{u}_j(\mathbf{r})$ of a classical field is a particular form of a field, which satisfies Maxwell's equations, and a set of which can serve as a mode basis. A common example is given by plane-wave modes propagating in vacuum,

$$\mathbf{u}_j(\mathbf{r}) = \mathbf{e}_j \exp\left[ i(k_{xj}x + k_{yj}y + k_{zj}z) \right], \tag{1}$$

where *j* is the mode index (label) and $\mathbf{e}_j$ are polarization vectors. In a classical description, the energy content of any mode may assume a continuous spectrum of values, proportional to the square of the field amplitude.

It is understood that modes of different frequencies can be added or superposed linearly with differing complex amplitudes $a_j$ to form the (real-valued) electric field, expressed mathematically as $\mathbf{E}(\mathbf{r},t) = \mathbf{E}^{(+)}(\mathbf{r},t) + cc$, where the 'positive-frequency part' of the complex field is represented by

$$\mathbf{E}^{(+)}(\mathbf{r},t) = \sum_{j=1}^{\infty} \mathcal{E} a_j e^{-i\omega_j t} \mathbf{u}_j(\mathbf{r}), \tag{2}$$

with a similar expression for the magnetic field. Bold-face fonts represent vector quantities and $\mathcal{E}$ is a scalar factor. The amount of energy 'occupying' a given mode is proportional to $|\mathcal{E} a_j|^2$. The mode's shape and propagation direction are contained in the forms of the $\mathbf{u}_j(\mathbf{r})$, which form a mutually orthogonal and complete set of functions. As seen by the time evolution $\exp(-i\omega_j t)$, the field in each mode undergoes single-frequency oscillations and can be described as a simple harmonic oscillator.

Each mode can be viewed as a separate subsystem, the totality of which form the overall field. When we discuss states of the overall field, in general we need to specify composite states involving the states of more than one mode. Such composite states can imply correlations between measurement outcomes on different modes.

While Eq.(2) is written as a discrete sum of modes as appropriate in a closed cavity, in unbounded free space the expression for the classical field becomes an integral over a continuum of frequencies. For a beam or pulse propagating in more-or-less a single direction, it is convenient to express it as the integral

$$\mathbf{E}^{(+)}(\mathbf{r},t) = \sum_{\sigma=1,2} \int_{-\infty}^{\infty} \frac{d\omega}{2\pi} \mathcal{E} a^{(\sigma)}(\omega) e^{-i\omega t} \mathbf{u}^{(\sigma)}(\omega,\mathbf{r}), \tag{3}$$

where $\sigma$ labels one of two polarization helicities in the case of circular polarization. In the simplest cases, such as in a waveguide or a well-collimated beam in free space, the spatial mode can be separated into transverse and longitudinal parts,

$$\mathbf{u}^{(\sigma)}(\omega,\mathbf{r}) = \mathbf{e}^{(\sigma)} w^{(\sigma)}(\omega,x,y) \exp(ik_z^{(\sigma)}(\omega) z), \tag{4}$$

where $\mathbf{e}^{(\sigma)}$ is a polarization vector, $w^{(\sigma)}(\omega,x,y)$ is the transverse part of the mode function, and the function $k_z^{(\sigma)}(\omega)$ is a dispersion relation (relationship between propagation constant and frequency).

In these equations, $\mathbf{E}(\mathbf{r},t)$ represents the vector-valued amplitude of the field. Of course, the physical field itself is distinct from its representation; the symbol $\mathbf{E}(\mathbf{r},t)$ is not the field itself, rather it is a *description* of the field, and many physicists take the field to be an actual element of physical reality. We follow this way of talking in this paper.

How do we describe the *state* of the field? A general definition of *state* can be said to be a description of everything that is known about the condition of a physical system at a certain time. In the simplest classical picture, we can have complete knowledge of the field, and the mathematical forms of $\mathbf{E}(\mathbf{r},t)$ and the magnetic field $B(\mathbf{r},t)$ give a full description of its state, that is, a specific configuration of the classical system. Often in optics problems the electric field dominates interactions with matter, such as detectors, and for freely propagating field the electric field is often sufficient for a complete description; so here we focus on describing the state of the electric field.

In general, however, we may possess only partial knowledge, in which case we describe the field's state using statistical means. For example, thermal light emitted by a blackbody is described as having field amplitudes that are random variables (or random processes) with zero mean value and Gaussian probability density. In principle, one could know the field values (that is, they have definite values at each instant), but in practice we don't. We call such a state a *statistical state*. To summarize:

**Definition**: A *classical state* is a description of the condition of a system, either a specific state representing completely possessed information or a statistical state representing partially possessed information.

In classical physics, a specific state is specified by a point in phase space, and along with the dynamical equations of motion specifies how the phase-space point evolves in time. On the other hand, a statistical state is described by a probability density function (pdf) giving joint probabilities of all possible values of system variables at all combinations of space-time points (for review see Mandel and Wolf, and Goodman [9,10]):

$$P_E(\mathbf{E}(\mathbf{r}_1,t_1),\mathbf{E}(\mathbf{r}_2,t_2),\mathbf{E}(\mathbf{r}_3,t_3),...). \tag{5}$$

With the classical mode decomposition, we can replace this pdf by a pdf for all the complex mode amplitudes,

$$P_a(a_1,a_2,a_3,...). \tag{6}$$

Any expectation values of quantities involving the field can, in principle, be calculated using either of these pdfs.

For a thermal-like classical state of a single mode, the pdf for the complex zero-mean random variable $a$ is

$$p_a(a) = \frac{1}{2\pi\sigma^2} e^{-|a|^2/2\sigma^2}, \tag{7}$$

where $2\sigma^2 = \langle |a|^2 \rangle$ is the variance of *a*. The corresponding pdf for the energy (proportional to $W = |a|^2$) in the mode is

$$p_W(W) = \frac{1}{\sqrt{2\pi\sigma^2}} e^{-W^2/2\sigma^2}. \tag{8}$$

When one speaks of 'mode' in optics, it is often assumed to be the spatial mode (as in a laser cavity). As in Eq.(3), one can always decompose the field in terms of products of spatial modes $\mathbf{u}^{(\sigma)}(\omega,\mathbf{r})$ and a multiplicative temporal factor $e^{-i\omega t}$. But in practice, such a monochromatic field would require an infinite time duration to fully define or measure. How can we realistically prepare and measure a single mode in the laboratory? A simple example is shown in **Fig.1**: open a small hole in a blackbody cavity for a time $T$, then spatially filter the emerging light with a distant, small pinhole, then pass it through a spectral filter with small transmitting bandwidth $\Delta v$ such that $\Delta v\, T \ll 1$, as described theoretically in [11]. The probability for energy content in this case is given by Eq.(8).

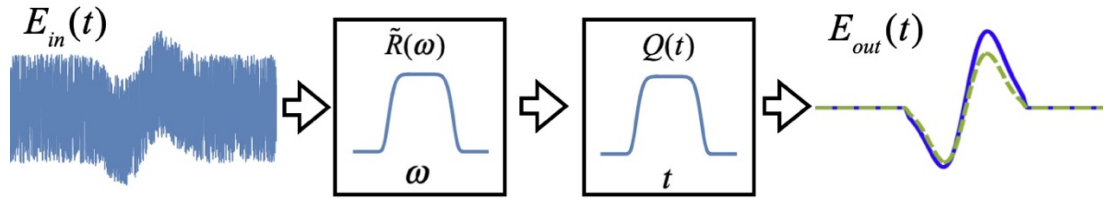

Fig. 1 A coherent signal pulse embedded in a noise background is filtered by passing through a sequence of a spectral filter $\tilde{R}(\omega)$ and a time gate $Q(t)$, resulting in a nearly coherent (single-temporal-mode) field. The envelope waveform (i.e., the field with the carrier wave removed) at the output illustrates both the ideal signal pulse (dashed curve) and the realistic simulated pulse (solid). Reproduced from [11].

Such a time-frequency filtering process selects one 'time-frequency mode,' also called a temporal-spectral mode, or temporal mode (TM) for short.[ii] [5, 12] In the following we will put an emphasis on temporal modes because they provide a mode basis for efficiently describing optical wave packets that are localized in space and time. Encoding and receiving information in such wave-packets form generally requires synchronizing a transmitter with the receiver.

Another important benefit of temporal modes is that they form a discrete set, rather than a continuous set as is the case for monochromatic modes. The discreteness makes it easier to distinguish one mode from another during a detection process. Their discreteness arises from the fact that, by definition, they are confined to a particular space-time region; that is, the boundary condition is that they go to zero at infinity in space and in time. This result is analogous mathematically to the quantization of spatial modes in a cavity or an optical fiber.

---

[ii] Note: Please don't confuse the abbreviation TM used here with the common terminology TM-mode, meaning transverse magnetic spatial mode.

To construct the TMs, consider a transformation from the monochromatic modes of Eq.(3) to the non-monochromatic temporal modes. We can choose any complete orthonormal set of spectral amplitude functions $\{f_j(\omega)\}$ that go to zero at frequencies far from a chosen central frequency $\omega_0$. By definition, they satisfy

$$\int \frac{d\omega}{2\pi} f_l^*(\omega) f_j(\omega) = \delta_{lj}$$
$$\sum_j f_j^*(\omega') f_j(\omega) = 2\pi \delta(\omega' - \omega) \qquad (9)$$

where $\delta_{lj}$ and $\delta(\omega)$ are the Kronecker and Dirac delta, respectively. We can use these functions to define a set of 'temporal-mode' amplitudes,

$$A_j = \int \frac{d\omega}{2\pi} f_j^*(\omega) a(\omega) \qquad (10)$$

Where $f_j(\omega)$ is the spectral amplitude defining such a mode labelled by $j$. Hereafter we drop the polarization label $\sigma$ for notational simplicity. The inverse relation is

$$a(\omega) = \sum_j f_j(\omega) A_j \qquad (11)$$

We see that the continuous (uncountably infinite) set of amplitudes $a(\omega)$ has been converted into a discrete (countably infinite) set $A_j$.

In terms of the TMs, the field in Eq.(3) is expressed as

$$\mathbf{E}^{(+)}(\mathbf{r},t) = \sum_j A_j \int_{-\infty}^{\infty} \frac{d\omega}{2\pi} \mathcal{E} f_j(\omega) e^{-i\omega t} \mathbf{e} w(\omega,x,y) \exp(ik_z(\omega)z)$$
$$\approx \mathcal{E} \mathbf{e} w(x,y) \sum_j A_j u_j(z,t) \qquad (12)$$

where, for simplicity, we assumed a common polarization vector $\mathbf{e}$ and made the approximation that the transverse mode function $w(\omega,x,y) \approx w(x,y)$ is independent of frequency, which is valid for reasonably narrow-band fields. Then the propagating temporal modes are defined as

$$u_j(z,t) = \int_{-\infty}^{\infty} \frac{d\omega}{2\pi} f_j(\omega) e^{-i\omega t} \exp(ik_z(\omega)z) \ . \qquad (13)$$

At position $z = 0$, the temporal mode equals the Fourier transform of the spectral amplitude:

$$u_j(0,t) = \int_{-\infty}^{\infty} \frac{d\omega}{2\pi} f_j(\omega) e^{-i\omega t} = \tilde{f}_j(t). \tag{14}$$

Several example sets of temporal modes are shown in **Fig.2**.

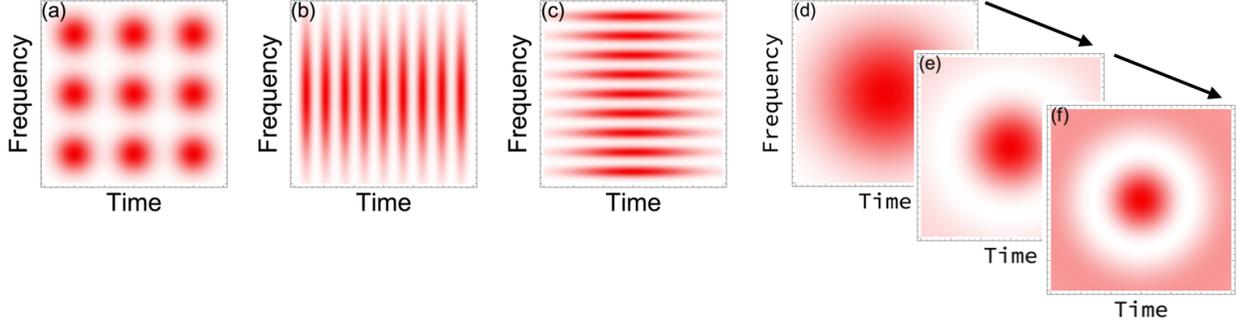

Fig. 2 Examples of sets of temporal modes, plotted as densities in a certain area of time-frequency phase (phasor) space. For a fixed scaling of the time and frequency axes, the modes may be equally broad in both variables as in a) or they may be narrow in time as in b) or narrow in frequency as in c). Such temporal modes may be Gaussian in form and are approximately orthogonal if their separations in time and frequency are large enough. As in d), e) and f), an alternative covering of the phase-space area can be accomplished using mode functions that cover the whole region and can be made strictly orthogonal in terms of coherent overlap integration as in Eq.(9).

In our time-frequency filtering example in **Fig. 1**, the strongly filtered field supports essentially only one TM,

$$\mathbf{E}^{(+)}(\mathbf{r},t) \approx \mathcal{E} A_{j=0}\, \mathbf{e} w(x,y) u_{j=0}(z,t). \tag{15}$$

Thus, the space and time behavior of the field is determined by the spatial-temporal mode $w(x,y)u_{j=0}(z,t)$, while the (classical) state is determined by the value of (or the statistical properties of) the mode amplitude $A_{j=0}$. The description in terms of temporal modes carries over directly to the quantum treatment of light.

Note that the 'incoherent' time-frequency filtering method described in **Fig. 1** is necessarily inefficient, in that to achieve a nearly single-temporal-mode field, the filtering needs to be so strong as to pass almost no light. Superior efficiency, approaching 100%, can be achieved using 'coherent' filtering with a scheme called a quantum pulse gate, as reviewed in [12] and utilized for noise filtering in [11].

## 3. Quantum Fields, Modes, States, and Photons

The fact that the classical EM field can be decomposed into a set of classical oscillators inspires us to seek a representation of the EM field as a collection of (bosonic) quantum harmonic oscillators. The creation and annihilation operators are labeled by the continuous frequency variable $\omega$ and are defined to obey the commutation relations [13, 14]

$$[\hat{a}(\omega),\hat{a}^{\dagger}(\omega')] = 2\pi\delta(\omega-\omega'). \tag{16}$$

The noncommutativity of operators for a given frequency embodies the essential difference between quantum and classical theory (and the nature of the physical systems being described). The energy eigenstates of each quantum oscillator are quantized: they occur at specific, identically-spaced values, $m\hbar\omega$ with $m$ a nonnegative integer. In the quantum theory, the mathematical representation of the field is given by the (Hilbert-space) operator (operators being indicated by carets),

$$\hat{\mathbf{E}}^{(+)}(\mathbf{r},t) = \sum_{\sigma=1,2} \int_{-\infty}^{\infty} \frac{d\omega}{2\pi} \mathcal{E}(\omega) \hat{a}^{(\sigma)}(\omega) e^{-i\omega t} \mathbf{u}^{(\sigma)}(\omega,\mathbf{r}). \tag{17}$$

The form of the commutator Eq.(16) along with energy quantization requires the scale factor to be frequency dependent, $\mathcal{E}(\omega) = (\hbar\omega/2\varepsilon_0 cn)^{1/2}$, where $\varepsilon_0$ and $c$ are the vacuum electric permittivity and speed of light and $n$ is the medium's refractive index at the frequency of interest.

For simplicity we neglect modal dispersion as occurs in a waveguide geometry, which is discussed in detail in [15]. This simplification allows us to drop any mode labels that refer to which waveguide mode is being considered.

The mode functions $\mathbf{u}^{(\sigma)}(\omega,\mathbf{r})$ are the same as in the classical theory and thus satisfy Maxwell's equations. As in the classical theory, a transformation from monochromatic modes to temporal (wave-packet) modes can be made using Eq.(13) for reasonably narrow-band fields. Then the field operator is

$$\hat{\mathbf{E}}^{(+)}(\mathbf{r},t) = \mathcal{E}\, \mathbf{e}\, w(x,y) \sum_j \hat{A}_j u_j(z,t), \tag{18}$$

where the scale factor for a center frequency $\omega_0$ is $\mathcal{E} = (\hbar\omega_0/2\varepsilon_0 cn)^{1/2}$ and $\hat{A}_j$ are annihilation operators for the state of light 'occupying' the temporal mode $u_j(z,t)$. They are given by

$$\hat{A}_j = \int \frac{d\omega}{2\pi} f_j^*(\omega) \hat{a}(\omega), \tag{19}$$

and from Eq.(16) it is easily shown that they satisfy the discrete, rather than continuous, commutation relation

$$[\hat{A}_i, \hat{A}_j^{\dagger}] = \delta_{ij}. \tag{20}$$

In free space (that is, no interactions) the time evolution, expressed in the Heisenberg Picture, is fully contained in the wave-packet propagation of the modes in Eq.(13).

(By the way, the assumption of narrow-band wave packets is not an essential step, but when this approximation is removed it turns out that the resulting wavepackets are not strictly orthogonal in space. This complication arises only for wave packets whose duration is less than around 10 fs, not usually the case in optical communications applications. See [16].)

Now we may ask, "What is the quantum field?" It is not $\hat{\mathbf{E}}^{(+)}(\mathbf{r},t)$, which is an operator that represents mathematically the annihilation of energy excitations in the field. The quantum field itself (in one meaningful way to view it) is a physical entity that is "out there" and is capable of carrying energy and momentum from one place to another.

How do we specify states of the quantum field?

**Definition**: A *quantum state* is a mathematical form used to determine the probabilities for particular outcomes of any possible measurement on a system, either as a pure state (representing maximal possessed information) or a mixed state (representing partial possessed information).

The most general case is the mixed state, expressed mathematically as a density operator,

$$\hat{\rho} = \sum_j P_j |\psi_j\rangle\langle\psi_j|, \tag{21}$$

where $P_j$ is the (classical) probability that the system is in the pure state $|\psi_j\rangle$. We say such a state is a statistical mixture of pure states. In the ideal limit, if all $P_j$ are known to be zero except a single one, say $P_0 = 1$, then we can describe the state simply by specifying the form of $|\psi_0\rangle$, which in some cases is described by a wavefunction of some systems variable or in other cases as a vector in an abstract linear vector space.

A starting point for the quantum state description of the field is the vacuum state, $|vac\rangle$, which carries no energy or momentum (at least not in a way that is detectable via absorption by an atom or a photodetector). The simplest non-vacuum state of the field is the single-photon state, and its generalization the *n*-photon (Fock) state, written in the temporal-mode formalism for mode $u_j(z,t)$ as

$$|1\rangle_j = \hat{A}_j^\dagger |vac\rangle = \int \frac{d\omega}{2\pi} f_j(\omega) \hat{a}^\dagger(\omega) |vac\rangle$$
$$|n\rangle_j = \frac{1}{\sqrt{n!}} \left(\hat{A}_j^\dagger\right)^n |vac\rangle \tag{22}$$

where $\hat{A}_j^\dagger$ is the creation operator for a given TM and *n* is the photon occupation number of a given TM of the field.

It is interesting that although an *n*-photon state of a particular temporal mode does contain a specific sharp number of photons, it does not contain a sharp value of energy, because the mode itself is constructed as a sum of modes having differing frequencies and each frequency mode has an unspecified number of photons (although they must sum to *n*). If the light is well collimated, quantum measurement of its energy content can be carried out using a spectrometer—disperse it on a diffraction grating followed by a dense array of photon-counting detectors. If the number of detectors is much greater than the number of photons *n*, and the detectors are 100% efficient, then exactly *n* detectors will register a detection event ('click'). Each detector will correspond to a resolved channel *l* with energy $\hbar\omega_l$. For $n=1$, the probability for a given detector to click is given by $|f_j(\omega)|^2$. In general, the pattern of detectors that click will indicate the total energy observed for that measurement trial.

To sum up, in the quantum theory of a collection of oscillators of different frequencies (in the temporal-mode formalism), what gets quantized is not total energy, but total excitation. You can have zero, one, or two excitations, but not half an excitation.

A general *pure state* of the field in a given temporal mode is expressed as

$$|\psi\rangle_j = \sum_{n=0}^{\infty} c_n |n\rangle_j = \sum_{n=0}^{\infty} c_n \frac{1}{\sqrt{n!}} \hat{A}_j^{\dagger n} |vac\rangle, \qquad (23)$$

where $|c_n|^2$ is the probability to observe *n* clicks in a detector array, as just described. The coherent state, with $c_n = \exp(-|\alpha|^2/2)\alpha^n/\sqrt{n!}$ for some complex amplitude $\alpha$, is the state from an ideal laser emitting a pulse in the temporal mode $u_j(z,t)$. Then the probabilities are given by the Poisson distribution $|c_n|^2 = \exp(-|\alpha|^2)|\alpha|^{2n}/n!$.

A *mixed state* of the EM field can be represented by a density operator for a given temporal mode,

$$\hat{\rho} = \sum_{n=0}^{\infty} p_n \frac{1}{\sqrt{n!}} \hat{A}_j^{\dagger n} |vac\rangle \langle vac| \frac{1}{\sqrt{n!}} \hat{A}_j^n, \qquad (24)$$

where $p_n$ is the probability to find *n* photons in mode *j* if measured. Interestingly, there is a different kind of mixed state, which has a definite number of photons (field excitations), but spread incoherently across several modes, for example, a single-photon mixed state is

$$\hat{\rho} = \sum_j P_j \hat{A}_j^{\dagger} |vac\rangle \langle vac| \hat{A}_j, \qquad (25)$$

where $P_j$ is the probability that the photon will be found in mode *j* if detected in a mode-selective manner.

## 4. Particles or Fields?

Now we can try to clarify, "What is a photon?" It is preferable not to think of the photon as a thing or a physical entity, rather as simply one of the names we use to specify states of the field. So, when we say, "The atom emitted a photon," we actually mean, "The atom lost energy, creating a single-photon state of the field." It is almost always safe (prudent) to replace "photon" by "single-photon state of the field." If we wish to have a more physically suggestive way to define a photon, we can say it is as a single 'excitation' of the quantum field.

Nevertheless, we note that a single-photon state can be thought of in two equivalent ways: as the state of a photon as a distinct entity, or as the state of a field. Consider such a state occupying a particular wave-packet mode $u_1(\mathbf{r},t)$ that is concentrated in a finite volume and traveling through space. Horizontal and vertical polarization states $|H\rangle, |V\rangle$ define a basis for describing the state of the photon. Say the photon's state is diagonal: $|D\rangle = (|H\rangle + |V\rangle)/\sqrt{2}$. Alternatively, we can write this photon state in terms of the state of the field, using modes that specify the field's spatial and polarization aspects denoted

$$\begin{aligned}\mathbf{u}_H(\mathbf{r}) &= \mathbf{e}_H u_1(\mathbf{r},t) \\ \mathbf{u}_V(\mathbf{r}) &= \mathbf{e}_V u_1(\mathbf{r},t)\end{aligned} \quad (26)$$

where $\mathbf{e}_H, \mathbf{e}_V$ are polarization vectors. Then we can express the single-photon diagonal-polarization state in terms of the occupation numbers of the $\mathbf{u}_H(\mathbf{r}), \mathbf{u}_V(\mathbf{r})$ modes,

$$\left(|1\rangle_H |0\rangle_V + |0\rangle_H |1\rangle_V\right)/\sqrt{2}. \quad (27)$$

Furthermore, we can choose a different mode basis to represent the same state. If we choose modes that define the field as diagonal and anti-diagonal polarized, $\mathbf{u}_D(\mathbf{r}) = (\mathbf{u}_H(\mathbf{r}) + \mathbf{u}_V(\mathbf{r}))/2^{1/2}$, $\mathbf{u}_A(\mathbf{r}) = (\mathbf{u}_H(\mathbf{r}) - \mathbf{u}_V(\mathbf{r}))/2^{1/2}$, then the same state is represented as $|1\rangle_D |0\rangle_A$. Thus, a transformation of the 'mode basis' from $\mathbf{u}_H(\mathbf{r}), \mathbf{u}_V(\mathbf{r})$ to $\mathbf{u}_D(\mathbf{r}), \mathbf{u}_A(\mathbf{r})$ corresponds to a change of the 'state basis' from $|H\rangle, |V\rangle$ to $|D\rangle, |A\rangle$.

The quantum theory of light can be constructed from either of two distinct starting points, as made clear by Dirac in his book, [17] where he wrote,

"The dynamical system consisting of an assembly of similar bosons is equivalent to the dynamical system consisting of a set of oscillators—the two systems are just the same system looked at from two different points of view."

Given the equivalence of the two points of view, why do many quantum physicists find it more fruitful to consider the field (that is, the collection of oscillators) as the fundamental physical entity, as we have done above. Steven Weinberg, a Nobel-winning quantum theorist, said, "Thus, the inhabitants of the universe were conceived to be a set of fields—an electron field, a proton field, an electromagnetic field—and particles were reduced to mere epiphenomena." The reasons for this choice are many and have been summarized in language suitable for the general reader in several accounts, including Hobson [18], Wilczek [19], and Raymer [20]. Here we offer a summary of such arguments.

1. Quantum fields respect relativity.

Frank Wilczek, also a Nobel-winning quantum theorist, writes, "The concept of locality, in the crude form that one can predict the behavior of nearby objects without reference to distant ones, is basic to scientific practice." Quantum field theory successfully describes all known phenomena (that it has been applied to) without invoking action at a distance, which would violate Einstein's relativity. [19]

2. Quantum fields evince identical particles.

Wilczek also writes, "Undoubtedly the single most profound fact about Nature that quantum field theory uniquely explains is the existence of different, yet indistinguishable, copies of elementary particles." It is known that the world is made of a limited number of particle types, and that any two members of the same type are identical. For example, any two electrons are identical, that is interchangeable. "We understand this as a consequence of the fact that both are excitations of the same underlying 'ur-stuff,' the electron field. The electron field is thus the primary reality," Wilczek says. And the same holds true for the electromagnetic field and its photon excitations. [19]

3. Quantum fields naturally account for changing numbers of particles.

Quantum field theory not only accounts for the creation and destruction of photons when atoms emit or absorb light, it also accounts for processes such as creation and destruction of electrons and positrons. Wilczek writes, "In this picture it is only the fields, and not the individual objects they create and destroy, that are permanent." [19]

4. Quantum fields give a clearer picture of wave-particle duality.

The electron-matter field is not an electron. Rather, an electron is an individual excitation of the electron-matter field, just as a photon is an individual excitation of the EM field. The quantum fields themselves behave in a wave-like manner and represent possible measurements to determine where the electron (or photon) is located. Therefore, it is not surprising that if one mistakenly believes an electron, for example, is a particle, meaningless questions can arise. For example, the question, which path did the electron take on its way to a detector, has no meaning. On the other hand, a quantum field permeates all of space; therefore, it exists within both paths. So, the proper statement is not that an electron sometimes behaves like a wave and sometimes like a particle. Rather one should say, the quantum field always behaves like a quantum field with its wave-like behaviors, and the electron is a manifestation of that field. It is best to replace the mysterious concept 'wave-particle duality' by the less mysterious concept 'quantum field-quantum particle duality.' (adapted from [20])

## 5. Superposition, Separability, and Entanglement

How is entanglement different from superposition? For a single quantum entity $|\psi\rangle_A + |\phi\rangle_A$ is a superposition state, wherein the "+" symbol represents the *superposition of possibilities* and can be read, "in superposition with." Here normalization factors have been dropped. For a pair of entities, $|\psi\rangle_A |\phi\rangle_B + |\phi\rangle_A |\psi\rangle_B$ is also a superposition state, but now involving a larger state space. A *nonseparable* state of two entities $A$ and $B$ is one that *cannot* be written as a product state, $|\psi\rangle_A |\psi'\rangle_B$. Fix to lower case (For simplicity we consider only pure states.)

It's worth unpacking what is meant by 'quantum entity.' In quantum photonics the entities are the field modes, and if they are correlated or entangled we need to specify a composite (joint) state in a higher-dimension state (Hilbert) space. Whether or not we call a nonseparable state entangled depends on the situation and to some extent the semantic preference of the user. According to some users, it is fair to call all nonseparable quantum states entangled. That would include the state of the two electrons in a helium atom. Other users will insist on reserving the use of 'entangled' for cases in which the two entities are not interacting (unlike two electrons in a helium atom) and can be measured independently in separated regions of space. The justification for this stricter definition is that in quantum information science entanglement is regarded as a resource for accomplishing tasks such as teleportation of a state across some distance. In this context, a powerful known fact is that the amount of such entanglement cannot be increased by any quantum operations in which each entity is transformed only locally, even in the presence of classical (non-quantum) communication between parties at the two locations. If one wishes to have distinct names to classify entanglement, one could say that nonseparability with spatial separation is called 'useful entanglement.'

For a two-mode nonseparable photonic state, can we transform to a new mode basis for which the state is separable? From the above discussion we see the answer is "No" if the entities are spatially separated and we restrict the basis changes to involve only local states of each entity. But if we carry out arbitrary global mode-basis change involving all relevant modes of the combined system, the answer is, "Yes, for all typically realizable states." For example, note that we can write the above-mentioned single-photon state (now with normalization) as

$$\frac{1}{\sqrt{2}}\left(|1\rangle_A |0\rangle_B + |0\rangle_A |1\rangle_B\right) = \frac{1}{\sqrt{2}}\left(\hat{A}^\dagger + \hat{B}^\dagger\right)|vac\rangle, \tag{28}$$

where $\hat{A}^\dagger, \hat{B}^\dagger$ are creation operators for two spatially orthogonal modes, as represented by Eq.(19). Consider the two-mode transformation of the spectral amplitudes,

$$\begin{pmatrix} f_A(\omega) \\ f_B(\omega) \end{pmatrix} \rightarrow \begin{pmatrix} f_C(\omega) \\ f_D(\omega) \end{pmatrix} = \frac{1}{\sqrt{2}}\begin{pmatrix} 1 & -1 \\ 1 & 1 \end{pmatrix}\begin{pmatrix} f_A(\omega) \\ f_B(\omega) \end{pmatrix}, \tag{29}$$

which is equivalent to the transformation of the mode functions on a 50/50 beam splitter,

$$\begin{pmatrix} u_A(z,t) \\ u_B(z,t) \end{pmatrix} \to \begin{pmatrix} u_C(z,t) \\ u_D(z,t) \end{pmatrix} = \frac{1}{\sqrt{2}} \begin{pmatrix} 1 & -1 \\ 1 & 1 \end{pmatrix} \begin{pmatrix} u_A(z,t) \\ u_B(z,t) \end{pmatrix}, \tag{30}$$

such that the operators transform as

$$\begin{pmatrix} \hat{A}^\dagger \\ \hat{B}^\dagger \end{pmatrix} \to \begin{pmatrix} \hat{C}^\dagger \\ \hat{D}^\dagger \end{pmatrix} = \frac{1}{\sqrt{2}} \begin{pmatrix} 1 & 1 \\ -1 & 1 \end{pmatrix} \begin{pmatrix} \hat{A}^\dagger \\ \hat{B}^\dagger \end{pmatrix}. \tag{31}$$

The state in Eq.(28) can then be written as

$$\frac{1}{\sqrt{2}}(\hat{A}^\dagger + \hat{B}^\dagger)|vac\rangle = \hat{C}^\dagger |vac\rangle = |1\rangle_C |0\rangle_D. \tag{32}$$

This is just a formal way to say that sending a single-photon state into a 50/50 beam splitter yields a mode-entangled state of the two emerging fields. The beam-splitter transformation is reversible, so we can send the mode-entangled state into the ports of a beam splitter and end up with a single-photon state in mode C (or in mode D if we adjust the phase). Thus, the state is nonseparable (entangled) in one mode basis but not in the other mode basis. Physically, undoing the entanglement (nonseparability) requires bringing the modes together to perform a 'global' transformation on them. As said above, the disentanglement cannot be accomplished by any 'local' transformations involving modes A and B separately.

Sperling et. Al. pointed out that there exist entangled (nonseparable) states that cannot be disentangled (made separable) by any unitary mode transformation, global or not. [7] At present, such states are not readily created using common techniques and have not played a role in quantum information science.

A very interesting point is that entanglement can exist with only one 'particle' present. If we consider two EM field modes, *A* and *B*, that occupy separate regions of space (e.g. two well-separated wave-packet modes $u_A(\mathbf{r},t)$ and $u_B(\mathbf{r},t)$), we could prepare each mode so it contains either zero or one photon's worth of quantum-field excitation, but there is only one photon in total. Label the state of the EM field in each mode by either (1) if it has one photon's worth of excitation, or (0) if it has none. The state $|1\rangle_A |0\rangle_B + |0\rangle_A |1\rangle_B$ refers to an entangled state of the fields in two modes. It may be helpful to recognize that modes represent different degrees of freedom of the EM field.

The state is not an entanglement of particle states. If you 'believe' in photons as particles, the state would be a mere superposition, $|A\rangle + |B\rangle$, not entangled in that context. The fact that two distinct EM modes can have entanglement even when there is only one photon shared between them suggests again that EM fields, not particles, are truly the physical entities. [6]

Note that there is an important difference between the joint state of the *A*, *B* modes being considered here and the state of the *H*, *V* modes in Eq.(27). The latter represents an excitation that is necessarily confined to a single spatial region, defined by the common spatial mode. Therefore, such a state is not considered to be the same kind of quantum resource as is the entangled state of two spatially separated modes. To make the point more mathematical, note that if we were to specify the latter state fully, we should indicate the state of all the relevant degrees of freedom of the field, including spatial location and polarization. Therefore, the state being considered in the present example is actually, assuming a common (*H*) polarization,

$$|1\rangle_{A,H}|0\rangle_{A,V}|0\rangle_{B,H}|0\rangle_{B,H} + |0\rangle_{A,H}|0\rangle_{A,V}|1\rangle_{B,H}|0\rangle_{B,H} \tag{33}$$

## 6. Mode Errors and State Errors

We turn to the question of what types of errors can occur when quantum information is being encoded in the states of light, such as in a quantum communication system.

We saw in the examples just given that a mode transformation can be thought of as a change of state, which is a Schrodinger-Picture way of thinking. Or we can think, in a Heisenberg-Picture way, that the global state has not changed, but only the state basis has been altered. Both ways of thinking are valid if we keep our pictures clear. Nevertheless, the distinction between modes and states can have practical consequences, or at least can allow one to categorize 'errors' that might occur in a quantum information science (QIS) scheme such as a quantum network.

Let's say we created a single-photon state as a superposition of two time-bin states with complex amplitudes $\alpha$ and $\beta$,

$$|\psi\rangle = \left(\alpha \hat{A}_1^\dagger + \beta \hat{A}_2^\dagger\right)|vac\rangle = \alpha|1\rangle_{u1}|0\rangle_{u2} + \beta|0\rangle_{u1}|1\rangle_{u2}, \tag{34}$$

where *u*1 and *u*2 refer to temporal modes $u_1(z,t)$ and $u_2(z,t)$ defined by Eq.(13) and taken here to be separated time-bin modes that are orthogonal to good approximation. We could encode a qubit using the two states

$$|\psi\rangle_1 = |1\rangle_{u1}|0\rangle_{u2} \; , \; |\psi\rangle_2 = |0\rangle_{u1}|1\rangle_{u2}, \tag{35}$$

or we could choose a transformed ("rotated") basis in which to encode the qubit,

$$\begin{aligned}|\psi\rangle_+ &= 2^{-1/2}\left(|1\rangle_{u1}|0\rangle_{u2} + |0\rangle_{u1}|1\rangle_{u2}\right)\\ |\psi\rangle_- &= 2^{-1/2}\left(|1\rangle_{u1}|0\rangle_{u2} - |0\rangle_{u1}|1\rangle_{u2}\right)\end{aligned}. \tag{36}$$

These four states are illustrated in Fig.3.

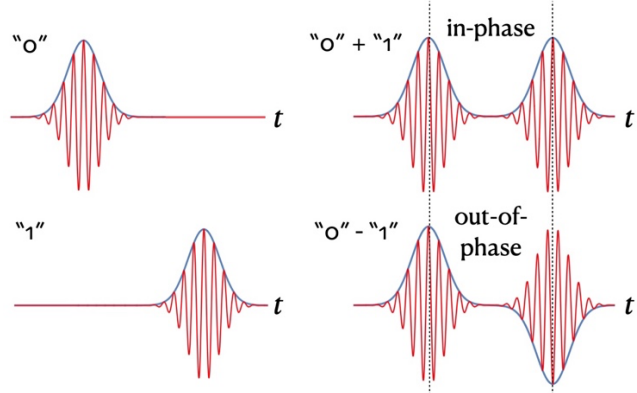

Fig. 3 A qubit can be encoded in two possible bases. On the left are two time-bin states that are nearly orthogonal, and on the right are two superposition states, which are nearly orthogonal by virtue of a relative phase shift of the components, and can equally well be used as a basis.

These two choices of bases are called mutually unbiased bases (MUBs). [21] If a qubit is created in any of the two states in one MUB, the probabilities for detecting the qubit in either of the states of the other MUB are equal and thus unbiased. Such pairs of MUBs play important roles in quantum key distribution (QKD). For a fun simulation of QKD see [22].

Given that we launch one of these four states into a QIP system such as a quantum network, two kinds of errors can occur—state errors and mode errors. Recall that we are thinking about states of fields, not of particles.

A *state error* occurs if coefficients of the state expansion (e.g. in Eq.(36)) are modified, while the forms of the modes (in this example time-bin or temporal modes) remain unchanged. For example, photons (field excitations) could be lost from the modes of interest by scattering or absorption, never to be recovered. Or photons could be added by leakage of light from other modes into the modes of interest. Another kind of state error is non-deterministic (random) dephasing between components of the state, for example,

$$|\psi\rangle_+ \to 2^{-1/2}\left(|1\rangle_{u1}|0\rangle_{u2} + e^{i\phi(t)}|0\rangle_{u1}|1\rangle_{u2}\right), \tag{37}$$

where $\phi(t)$ is an unknown, uncontrollable phase. Such a change would drive the pure state into a mixed state, from which the original pure state cannot be recovered.

In contrast, a *mode error* occurs if the forms of modes become altered by some physical process, while the state remains unchanged. Linear dispersion in a fiber, which is deterministic, can often be reversed or otherwise compensated by a physical process such as a prism pair. Alternatively, one can simply redefine the modes of interest to be those that exist following the predictable effects of dispersion. Notice that dispersion is already accounted for in the definition of the temporal modes $u_j(z,t)$ in Eq.(13) through the dispersion relation $k_z(\omega)$. Another kind of deterministic mode change occurs simply by time delay, either of the two-pulse wave packet as a whole or a change of the time delay between the time bins being used to encode the qubit. As with dispersion, such changes can be compensated or accounted for theoretically. These examples point out the importance of knowing any changes of the modes during propagation, as well as the need for synchronization in most designs for a quantum network (although important progress has been made toward using single-atom quantum memories to remove the need for synchronization [23]).

Other kinds of deterministic (unitary) mode changes can also occur, such as linear mixing of the modes of interest with other modes in processes analogous to linear beam splitting. Such process might be reversible if all the involved modes can be controlled.

Interestingly, the state change illustrated in Eq.(37) can be interpreted instead as a mode change, in particular,

$$u_1(z,t) \to u_1(z,t), \quad u_2(z,t) \to e^{-i\phi(t)} u_2(z,t), \tag{38}$$

that is, a random dephasing between temporal-mode components of the state. That is, in this case the error may be viewed as either a mode or state error since the phase change can be absorbed into either the state coefficients or the mode functions.

## 7. Photon Wave Function

Finally, we discuss the question – if we for pedagogical reasons want to depart from the approach taken until now in this paper, which treats fields, not particles, as the fundamental quantities, how far can we go? Let's first remark that often one hears talk about 'which path' a photon might take in, for example, a double-slit experiment. Such a question presupposes that a photon is a particle-like entity that has a trajectory, or at least a set of possible trajectories. In contrast, in the field-theoretic approach, such a question never needs to be asked because the field fills all of space, so it makes no sense to ask which path it takes.

Possible answers to the question depend on how strict one wants to be in defining what a particle is. If one simply conceives of particles as discrete but otherwise abstract entities that carry energy and momentum, then there seems to be no problem defining a wave function that describes its properties and dynamics. However, if one insists that a particle be an entity that can be localized *to a point* in space, then complications arise.

The question therefore leads us to the intriguing possibility of defining a wave function for a photon—a topic that was explored in the early days of quantum theory and more recently and fruitfully by Iwo Białynicki-Birula, whom this paper is meant to honor. [1] For a wave function for a photon in coordinate space, there should be a corresponding Schrodinger equation that it satisfies. But what is this wave function and what is its Schrodinger equation? The simple answer is that the photon wave function is a wave packet that satisfies Maxwell's equations. Therefore, the single-photon wave function could be given by a monochromatic mode such as $\mathbf{u}^{(\sigma)}(\omega,\mathbf{r})$ in Eq.(3) or superpositions forming a temporal mode such as $u_j(z,t)$ in Eq.(13), and its Schrodinger equation has the form of Maxwell's equations. Let's explore this statement further.

As developed in [1, 2] and reviewed in [16], a compact way to write classical Maxwell's equations is to combine the (real) electric and magnetic fields, $\mathbf{E}(\mathbf{r},t)$, $\mathbf{B}(\mathbf{r},t)$, into a single complex field, called the Riemann-Silberstein (RS) vector field,

$$\Psi_\sigma(\mathbf{r},t) = \left(\frac{\varepsilon_0}{2}\right)^{1/2} \left(\mathbf{E}(\mathbf{r},t) + i\sigma c \mathbf{B}(\mathbf{r},t)\right), \tag{39}$$

where $\sigma = \pm 1$ describes the field's helicity (circular polarization). $\Psi_\sigma(\mathbf{r},t)$ represents two fields, one for each value of $\sigma$, and, if one wishes, it can be combined into a single two-entry entity $\{\Psi_{+1}, \Psi_{-1}\}$ (analogous to a Dirac spinor). For a single helicity, the free-space Maxwell's equations with no charges or currents present can be written as a vector cross product,

$$i\frac{\partial}{\partial t}\Psi_\sigma(\mathbf{r},t) = c\sigma \nabla \times \Psi_\sigma(\mathbf{r},t) \quad (\sigma = \pm 1). \tag{40}$$

We can think of this form as a Schrodinger equation,

$$i\hbar\frac{\partial}{\partial t}\Psi_\sigma(\mathbf{r},t) = H\Psi_\sigma(\mathbf{r},t), \tag{41}$$

where the Planck constant $\hbar$ has been inserted on both sides and the (Maxwell) Hamiltonian operator is defined as $H = \hbar c\sigma \nabla \times$.

It's interesting that Eq.(41) is a wave equation for the electromagnetic field that is first-order in the time derivative, whereas in classical optics we usually think of a wave equation that is second order. The first-order wave equation here is simply an alternate way to write the two Maxwell's equations together and it embodies more information than the familiar second-order wave equation. (Historically, the desire for a first-order-in-time wave equation is what drove Dirac to formulate his famous relativistic Schrodinger wave equation for the electron.)

One can fruitfully consider that Eq.(41) is the Schrodinger equation for a single photon (whether one regards the photon as a particle or as an excitation of the quantized EM field). This statement can be 'tested' by performing a so-called second quantization of the photon wave function to construct a quantum field theory that permits more than a single excitation. That is, replace the classical function by an operator, $\Psi_\sigma(\mathbf{r},t) \to \hat{\Psi}_\sigma(\mathbf{r},t)$,

$$\Psi_\sigma(\mathbf{r},t) \to \hat{\Psi}_\sigma(\mathbf{r},t) = \sum_j \hat{b}_j^{(\sigma)} \psi_j^{(\sigma)}(\mathbf{r},t) + \sum_j \hat{b}_j^{(\sigma)\dagger} \psi_j^{(\sigma)*}(\mathbf{r},t), \tag{42}$$

where the $\psi_j^{(\sigma)}(\mathbf{r},t)$ are (vector) mode functions and the creation and annihilation operators satisfy $[\hat{b}_j^{(\sigma)}, \hat{b}_k^{(\sigma')\dagger}] = \delta_{jk}\delta_{\sigma\sigma'}$. Hereafter for concreteness we consider a single helicity, $\sigma = +1$.

We identify this field operator as the complex sum of electric and magnetic field operators (indicated by carets),

$$\hat{\Psi}_{+1}(\mathbf{r},t) = \left(\frac{\varepsilon_0}{2}\right)^{1/2} \left(\hat{\mathbf{E}}(\mathbf{r},t) + ic\hat{\mathbf{B}}(\mathbf{r},t)\right). \tag{43}$$

Then we find for the electric field part, using

$$(2\varepsilon_0)^{-1/2}\left(\hat{\Psi}_{+1}(\mathbf{r},t) + \hat{\Psi}_{+1}^{\dagger}(\mathbf{r},t)\right) = \hat{\mathbf{E}}(\mathbf{r},t), \tag{44}$$

and defining positive and negative-frequency parts, $\hat{\mathbf{E}}(\mathbf{r},t) = \hat{\mathbf{E}}^{(+)}(\mathbf{r},t) + \hat{\mathbf{E}}^{(-)}(\mathbf{r},t)$, where

$$\begin{aligned}\hat{\mathbf{E}}^{(+)}(\mathbf{r},t) &= (2\varepsilon_0)^{-1/2} \sum_j \hat{b}_j \,\boldsymbol{\psi}_j(\mathbf{r},t) \\ \hat{\mathbf{E}}^{(-)}(\mathbf{r},t) &= (2\varepsilon_0)^{-1/2} \sum_j \hat{b}_j^{\dagger} \,\boldsymbol{\psi}_j^*(\mathbf{r},t)\end{aligned}, \tag{45}$$

and we dropped the $\sigma$ label for simplicity. Comparing with Eq.(18) and identifying the operators by $\hat{b}_j = \hat{A}_j$ and photon wave functions by $(2\varepsilon_0)^{-1/2}\boldsymbol{\psi}_j(\mathbf{r},t) = \mathcal{E}\,w(x,y)\mathbf{e}_j u_j(z,t)$, we see that the second quantization procedure leads directly to the spatial-temporal-mode formalism of quantum optics.

As long as we restrict our considerations to reasonably narrow-band fields in each mode, where the frequencies are near a central carrier frequency $\omega \approx \omega_0$, and the bandwidth is much smaller than $\omega_0$, we have $\mathcal{E}(\omega) \approx \mathcal{E} = (\hbar\omega_0/2\varepsilon_0 cn)^{1/2}$. Then the single-photon wave functions corresponding to different annihilation operators are orthogonal to good approximation, in the sense that integrating over all space yields

$$\int d^3r\,\boldsymbol{\psi}_j^*(\mathbf{r},t)\cdot\boldsymbol{\psi}_k(\mathbf{r},t) = \delta_{jk}. \tag{46}$$

A subtlety arises when considering exotic ultra-broadband photons with bandwidth comparable to (say 50% or greater than) the carrier frequency. Then a more careful analysis starting from Eq.(17) with the frequency dependence retained in $\mathcal{E}(\omega)$ shows that the photon wave functions corresponding to different annihilation operators cannot be strictly orthogonal in space. [16] This non-orthogonality means that, strictly speaking, a single-photon state cannot be localized to a point in space; that is there is no local spatial probability to find the photon (thinking of a particle) at any particular point. In general, it is more accurate to say that the modulus-squared of a photon wave function $|\boldsymbol{\psi}_j(\mathbf{r},t)|^2$ describes the spatial distribution of probabilities to detect the photon's *energy* concentrated around different locations $\mathbf{r}$, rather than to find the photon (as if it were a particle) at a specified point location. The photon, viewed as a particle or as a state of the field, always remains 'spread out' within a region with minimum volume equal to a cubic wavelength.

It should be mentioned that the same complication arises in the temporal-modes formalism when considering ultra-broadband temporal modes, because it is mathematically equivalent to the second-quantized photon-wave-function formalism. In practice, such details have not (yet) been found to have significant consequences in quantum information science, where ultra-broadband photons are not typically employed.

If there are two field excitations (photons) present, the concept of a two-photon wave function becomes relevant. Its modulus-squared gives the probability for finding the energies of the two photons concentrated around locations $\mathbf{r}_1$ and $\mathbf{r}_2$. By analogy with a two-electron wave function, such a function has been defined as (suppressing the polarization label for simplicity) [16]

$$\Psi(\mathbf{r}_1,\mathbf{r}_2,t) = \sum_{j,k} C_{j,k}\, \psi_j(\mathbf{r}_1,t) \otimes \varphi_j(\mathbf{r}_2,t), \tag{47}$$

where $\psi_j(\mathbf{r}_1,t)$ and $\varphi_j(\mathbf{r}_2,t)$ are single-photon wave functions and the product is a vector direct product. $C_{j,k}$ can be chosen arbitrarily as long as it ensures the symmetry properties of a two-boson state, namely $\Psi(\mathbf{r}_2,\mathbf{r}_1,t) = \Psi(\mathbf{r}_1,\mathbf{r}_2,t)$. Two-photon wave functions expressed in spatial coordinates are not often used in quantum optics theory because the quantized field method is nearly always more direct and convenient, especially in scenarios where the number of photons changes in time. A caveat is that the two-photon wave function is equivalent to the so-called *two-photon detection amplitude*, which arises naturally in standard quantum optics theory when considering joint detection of two-photon states. [24, 9] Such a formalism arises naturally from standard quantum optics theory when analyzing optical detection.

Thus, we see that a photon-as-particle viewpoint can be formulated and used if one is careful to understand its limitations and its relation to standard quantum optics theory, which is based on quantization of the EM field. Further developments have included treating the case when light interacts with matter; then the one- or two-photon wave-function approach has to be modified, as done, for example, by Saldanha and Monken and by Keller. [25, 26]

When considering fields with more than two excitations (photons), the wave-function picture quickly becomes inconvenient and overly cumbersome (it becomes a many-body quantum theory), again giving credence to the preference among theorists to stick with the quantized-field approach. [16]

## 8. Mode Interference and State Interference

Electromagnetic modes satisfy Maxwell's equations, and therefore optical interference is built into the quantum theory from the start. The mode transformation in Eq.(30) is an example of interference. What is perhaps confusing is that modes interfere (classically) and quantum-state amplitudes can also interfere (quantumly). For a single-photon state you can think of 'classical' mode interference (on a beam splitter, say) as a quantum-change of state or a change of mode basis, as in Eq.(30). For multi-photon states, or states with indefinite number of photons, the situation is more subtle.

As mentioned earlier, a coherent state of a single monochromatic mode is expressed as

$$|\alpha\rangle = e^{-|\alpha|^2/2} \sum_{n=0}^{\infty} \frac{\alpha^n}{n!} \hat{a}^{\dagger n} |vac\rangle = e^{-|\alpha|^2/2} \sum_{n=0}^{\infty} \frac{\alpha^n}{\sqrt{n!}} |n\rangle, \quad (48)$$

and is found (upon measurement) to contain $n$ photons with probability $e^{-|\alpha|^2} |\alpha|^{2n}/n!$, a Poisson distribution. If the coherent-state field passes through a phase-shifting element such as a piece of transparent glass, it picks up a phase shift $\theta$, which manifests in the coherent state as [27]

$$|\alpha\rangle = e^{-|\alpha|^2/2} \sum_{n=0}^{\infty} \frac{[\alpha e^{-i\theta}]^n}{\sqrt{n!}} |n\rangle = e^{-|\alpha|^2/2} \sum_{n=0}^{\infty} \frac{\alpha^n e^{-in\theta}}{\sqrt{n!}} |n\rangle, \quad (49)$$

that is, the $n$-photon component is phase shifted $n$ times more than the one-photon component, keeping the state still a coherent state. (This phase shift is simply the Schrodinger time-evolution factor $\exp(-in\hbar\omega t/\hbar)$, where $n\hbar\omega$ is the energy of the state component.) In this case, then, the 'quantum phases' of each state component are related simply to the 'classical phase' shift $\theta$. In this way, we can say there is "One phase to rule them all," with apologies to J.R.R. Tolkien. Such is not the case for more general (non-coherent) states of the field, where in general the state components $\exp(-i\theta_n)|n\rangle$ can have arbitrary values of their quantum phases $\theta_n$, depending on the means of their generation. This fact, along with the facts that quantum interference of coherent states mimics perfectly classical interference of fields and that photoelectron statistics (Poisson) are the same as in a semiclassical model of detection, [9, 28] is one reason that coherent states are considered to be the 'most classical' states possible.

A single-photon state also behaves classically in propagation in the sense that its wavefunction follows the classical Maxwell's equations, as pointed out in the previous section. The quantum nature of this state becomes apparent only when detection occurs; the photon is "found" to be localized (within a cubic wavelength) at only one detector, not simultaneously at two or more. Yet the average rate of detection events upon repeated trials yields results identical to those expected from a classical theory treatment augmented by detection statistics. For this reason, single-photon optics experiments are sometimes thought not to be 'classical enough' to bring out uniquely quantum aspects of nature. Correlations of detection events are more revealing, and such experiments require detection of at least two photons.

Paul Dirac famously tried to sum up the situation by saying, "Each photon then interferes only with itself." [8] Later it became clear that nature is not that simple. A state of the field in which two excitations are present, possibly in different modes, is called a 'biphoton.' The well-known Hong-Ou-Mandel (HOM) two-photon interference is illustrated in Fig. 4 and is understood by considering two single-photon states impinging on two separate input sides of a 50/50 beam splitter, using $\hat{C}^\dagger |n\rangle = (n+1)^{1/2} |n+1\rangle$,

$$|\Psi\rangle = |1\rangle_A |1\rangle_B = \hat{B}^\dagger \hat{A}^\dagger |vac\rangle = \frac{1}{\sqrt{2}}\left(\hat{C}^\dagger + \hat{D}^\dagger\right)\frac{1}{\sqrt{2}}\left(\hat{C}^\dagger - \hat{D}^\dagger\right)|vac\rangle$$
$$= \frac{1}{2}\left(\hat{C}^{\dagger 2} - \hat{D}^{\dagger 2}\right)|vac\rangle = \frac{1}{\sqrt{2}}\left(|2\rangle_C |0\rangle_D - |0\rangle_C |2\rangle_D\right) \quad . \tag{50}$$

That is, both photons are detected in either the C mode or in the D mode, with 50% probability; we never see coincidence events of a detector placed in the outgoing C mode with a detector in the D mode. [29] This phenomenon is a strictly quantum one and does not occur in classical electromagnetic theory. (Although there are classical mimics of this effect, the coincidence probability cannot go to zero in such examples.) Therefore, for two-photon interference, we can say, "Each biphoton interferes only with itself." [30, 31]

The general statement might be stated best as, "Upon detection, each quantum state component interferes only with itself, and only if they occupy the same mode." One might wonder, then, how to understand the common situation that occurs when two fields of different carrier frequencies come together and create intensity 'beats' at the difference frequency. It might seem to contradict the idea that modes interfere only with themselves. But it has to be noted that when time-resolved detection is performed in order to observe the beats, a mode projection takes place: the monochromatic modes of distinct frequencies are both projected onto a set of common temporal modes (fields confined within a certain short interval). These temporal modes then interfere with themselves. Such issues were understood as long ago as 1969. [32]

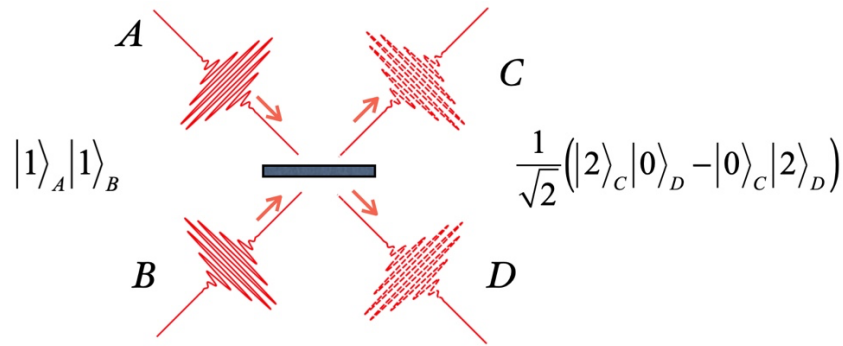

Fig.4 A separable, nonentangled biphoton state enters the input paths of a 50/50 beam splitter. The result is an entangled biphoton state in which both photons appear in one or the other output path. Note that, because single-photon states don't carry phase information per se, the introduction of a phase shift in either path before the beam splitter will not affect the two-photon interference outcome.

# 9. Conclusions

In summary: A classical field can be viewed as a physical entity that fills all of space and can transport energy and momentum in the form of continuous wave-like excitations. Similarly, a quantum field can be viewed as a physical entity that fills all of space and can transport energy and momentum only in the form of discrete excitations; these excitations exhibit both wave-like and particle-like behaviors. A photon is the label we give to a state representing a single excitation of the EM field. In most cases, it is safer to say, "a single-photon state of the field" rather than simply "a photon."

A convenient way to represent the state of the field is to decompose the field into a weighted sum of mode functions. In classical theory the weighting coefficients can take on continuous values, representing a continuum of possible energies. In quantum theory the weighting coefficients are operators, representing quantization (discreteness) of possible energies.

If we are careful, we can describe states in the field picture or in the particle picture, a viewpoint discussed in detail by Iwo Białynicki-Birula. We like to imagine that a given mode is like a container into which we can put any field state. For the special case of the single-photon state, "mode" in the field picture means the same as "state" in the particle picture. In that case, a single photon distributed coherently between distinct modes represents an entangled state, although such entanglement cannot be verified experimentally by simply detecting the photon in one mode or the other. The modes must interact with other physical entities such as separate atoms.

When more than one photon is present, we recognize that different modes can be put into any one of many possible combined field states—separable, correlated, or entangled. A biphoton is the name given to a state representing a double excitation of the EM field, whether the excitations are in the same or different modes. Modes of the field are essentially classical constructs and satisfy Maxwell's equations. As such they can interfere 'classically.' Quantum states can interfere 'quantumly,' as in the HOM effect. These two kinds of interference create a rich structure for quantum photonics.

In the context of optics and photonics, the distinction or boundary between quantum and classical is somewhat murky, although useful operational definitions have been developed.  Often, we define 'classical' to mean that detection statistics can be predicted correctly by a theory in which light is treated as a classical EM wave (although it may be random, stochastic) and the detectors (photo-emissive detectors - photo diode, photomultiplier) are treated by quantum theory. By this 'semiclassical' definition, coherent states and any mixture of them, such as thermal (e.g. blackbody) states, are considered classical. [9] Single-photon and biphoton states are then considered quantum. Still, there are cases where coherent states can be used to implement an intrinsically quantum task, such as QKD. There, a highly attenuated laser pulse can be engineered to be in one of several possible weak coherent states and the quantum behavior occurs upon detection. Any intermediate measurement of the state (such as by an eavesdropper) will necessarily cause a disturbance of the state and thus be detectable. The fundamentally lowest disturbance is dictated by quantum principles related to the Heisenberg Uncertainty Principle, and thus even though the state of light is considered 'classical,' the security of communication can be assured by the quantum physics of measurement. [33]

Harnessing the various degrees of freedom of the optical field, including the temporal-spectral one, can provide novel means to encoding and manipulating quantum information and is thus an ongoing topic of research. [12]

**Acknowledgements**

We thank Brian Smith for many discussions over the years about the contents of this paper. And MR thanks Iwo Białynicki-Birula for many helpful discussions about photon wave functions. This work was supported by the Engineering Research Centers Program of the National Science Foundation under Grant #1941583 to the NSF-ERC Center for Quantum Networks.